\documentclass[aps,prc,floatfix,preprint]{revtex4}

\preprint{JLAB-THY-11-1454}

\usepackage{amsmath}
\usepackage{graphicx}
\usepackage{bm}
\DeclareGraphicsExtensions{.ps}

\newcommand{\sW}{\sin^2\theta_W}

\newcommand{\fref}[1]{Fig.~\ref{fig:#1}}

\begin{document}

\title{Impact of PDF uncertainties at large $x$		\\
	on heavy boson production}
\author{L. T. Brady$^{1,2}$, A. Accardi$^{1,3}$,
	W. Melnitchouk$^1$, J. F. Owens$^4$}
\affiliation{
	$^1$Jefferson Lab, Newport News, Virginia 23606		\\
	$^2$Harvey Mudd College, Claremont, California 91711	\\
	$^3$Hampton University, Hampton, Virginia 23668		\\
	$^4$Florida State University, Tallahassee, Florida 32306}

% \date{\today}

\begin{abstract}
We explore the sensitivity of $W$ and $Z$ boson production in hadronic
collisions to uncertainties in parton distribution functions (PDFs) at
large $x$ arising from uncertainties in nuclear corrections when using
deuterium data in global QCD fits.  The $W$ and $Z$ differential cross
sections show increasing influence of nuclear corrections at high boson
rapidities, particularly for the $d$ quark, which is diluted somewhat in
the decay lepton rapidity distributions.
The effects of PDF uncertainties on heavy $W'$ and $Z'$ bosons beyond
the Standard Model become progressively more important for larger
boson masses or rapidities, both in $pp$ collisions at the LHC and
in $p\overline p$ scattering at the Tevatron.
\end{abstract}

\maketitle

%%%%%%%%%%%%%%%%%%%%%%%%%%%%%%%%%%%%%%%%%%%%%%%%%%%%%%%%%%%%%%%%%%%%%%%%%
\section{Introduction}
\label{sec:intro}

The discovery and determination of properties of new particles beyond
the Standard Model at high-energy colliders depends on accurate
knowledge of parton distribution functions (PDFs) of the hadrons
involved in the collisions.  With the Large Hadron Collider (LHC) at
CERN now taking data at unprecedentedly high energies, the effort to
control backgrounds in searches for the Higgs boson and other putative
particles is taking on paramount urgency.  Recently the dependence of
the Higgs boson cross section for the dominant gluon--gluon fusion
channel on PDFs has been the cause of some debate \cite{Alekhin11,
Thorne11, Baglio11}, highlighting the need for a careful determination
of strong interaction inputs such as gluon distributions, the strong
coupling constant $\alpha_s$, and higher-order radiative corrections.

At lower energies, the importance of PDF uncertainties has also been
discussed recently in fixed-target experiments, particularly for the
$d$ quark distribution in the region of large parton momentum fractions
$x$ ($x \gtrsim 0.5$) \cite{CTEQX,CJ11}.  Because both proton and
deuterium deep-inelastic scattering (DIS) data are required to
constrain the $d$ quark PDF, uncertainties in the nuclear corrections
in the deuteron at large $x$ translate into significant and growing
uncertainties on the $d/u$ ratio as $x \to 1$.  Through $Q^2$ evolution,
this can impact cross section calculations at smaller $x$ and larger
$Q^2$ \cite{Kuhlmann00}, especially in regions where the rapidity is
large.

In addition to Higgs boson cross sections, other processes studied
at the LHC or the Tevatron at Fermilab that may be sensitive to PDF
uncertainties include the production of heavy $W'$ and $Z'$ bosons
associated with additional SU(2)$\times$U(1) gauge groups.  These are
predicted in various extensions of the Standard Model, such as the
SO(10) and E$_6$ grand unified theories, or supersymmetric models,
some with $W'$ and $Z'$ boson masses at the TeV scale (for reviews
see, {\it e.g.}, Refs.~\cite{Rev:Hewett89, Rev:Leike99, Rev:Rizzo06,
Rev:Langacker09}).  Their production cross sections at large rapidities
will involve products of PDFs evaluated with one value of $x$ small
and the other large, thereby exposing them to uncertainties in PDFs
at large $x$, particularly near the kinematic limits.

In this paper we explore the sensitivity of the weak boson production
cross sections to uncertainties in PDFs at large $x$.  For our numerical
estimates we use the PDFs from the recent CTEQ-Jefferson Lab (CJ)
next-to-leading order (NLO) global analysis \cite{CJ11} of proton and
deuteron data, which quantified the model dependence of the nuclear
corrections in the deuteron and the resulting effects on the PDFs.
In Sec.~\ref{sec:du} we briefly review the CJ analysis and the origin
of the PDF uncertainties, before examining their impact in 
Sec.~\ref{sec:WZ} on the physical $W^\pm$ and $Z$ boson cross sections,
and $W$ and lepton charge asymmetries.
The effects of large-$x$ PDF uncertainties on production rates of
heavy $W'$ and $Z'$ bosons in $pp$ and $p\bar p$ collisions at the
LHC and Tevatron, respectively, are studied in Sec.~\ref{sec:prime}
as a function of the boson mass, and limits placed on the accuracy
with which cross sections for bosons of a given mass can currently
be determined.
(Even though data taking has now been completed at the Tevatron,
considerable quantities of data remain to be analyzed.)
Finally, some concluding remarks are made in Sec.~\ref{sec:conc}.

%%%%%%%%%%%%%%%%%%%%%%%%%%%%%%%%%%%%%%%%%%%%%%%%%%%%%%%%%%%%%%%%%%%%%%%%%
\section{PDF uncertainties at large \large $\bm{x}$ \normalsize}
\label{sec:du}

The CJ analysis \cite{CJ11} was a dedicated global NLO fit of proton
and deuteron DIS and other high-energy scattering data, which critically
examined the effects on PDFs of nuclear corrections in the deuteron
$F_2$ structure function.  Nuclear corrections were estimated using
a smearing function computed within the weak binding approximation
\cite{KP06,Kahn09}, taking into account nuclear binding and Fermi motion,
as well as a range of models describing the possible modification of
the nucleon structure function off-shell \cite{KP06,MSTprd,MSTplb}.
Several deuteron wave functions were considered, based on high-precision
nucleon--nucleon potentials, including the nonrelativistic CD-Bonn
\cite{CDBonn} and AV18 \cite{AV18} wave functions, and the relativistic
WJC-1 and WJC-2 wave functions \cite{WJC}, as well as the older Paris
\cite{Paris} wave function for reference.  The off-shell corrections
were estimated from a relativistic quark spectator model \cite{MSTplb},
and from a phenomenological model proposed by Kulagin \& Petti
\cite{KP06} but modified for the specific case of the deuteron
(see Ref.~\cite{CJ11} for details).

\begin{figure}[ht]
% \rotatebox{-90}{\includegraphics[width=6cm]{plots/du.eps}}
\rotatebox{-90}{\includegraphics[width=6cm]{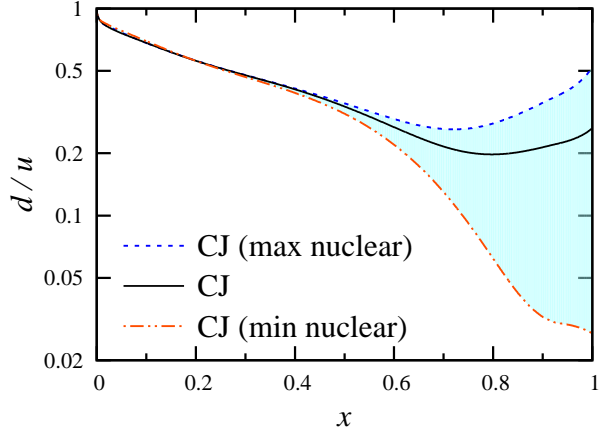}}
\caption{Ratio of $d$ to $u$ quark distributions for the CJ PDFs
	\cite{CJ11} at $Q^2 = 10$~GeV$^2$.  The shaded band illustrates
	the uncertainty range between maximum (blue dashed) and
	minimum (red dot-dashed) nuclear corrections in the deuteron.}
\label{fig:du}
\end{figure}

Combinations of deuteron wave functions and off-shell models giving     
the smallest and largest nuclear effects were identified, and used to
define the range of the nuclear corrections from the minimum (WJC-1
wave function and no off-shell corrections) to the maximum (CD-Bonn
wave function and largest off-shell corrections) nuclear corrections.
The central values, which are used as a reference, were obtained using   
the AV18 wave function and an intermediate off-shell correction.
The resulting fitted $d/u$ quark distribution ratio is shown in
Fig.~\ref{fig:du} for the full range of nuclear uncertainties
determined in Ref.~\cite{CJ11} (see also Refs.~\cite{HAMP11,ARM11}).
While the $u$ quark distribution is relatively well constrained
by proton structure function data for all values of $x$, the $d$
quark PDF has large uncertainties beyond $x \approx 0.5$.
Significant uncertainties appear also in the gluon PDF in the
region of $x$ not directly constrained by data, as well as in the
$\bar u$ and $\bar d$ distributions due to correlations with the
$d$ quark, induced in the global fits by moderate-$x$ jet and
dilepton production data, respectively.

Several experiments to directly measure $d/u$ up to $x \approx 0.8$
are planned at the 12~GeV energy upgraded Jefferson Lab in the near
future \cite{BoNuS12, MARATHON, SOLID}, which it is hoped will reduce
the uncertainties significantly.  
These include the ``MARATHON'' \cite{MARATHON} experiment, which aims
to extract $F_2^n/F_2^p$ from a measurement of the $F_2$ structure
functions of tritium and $^3$He with cancellation of nuclear effects
to the $\approx 1\%$ level \cite{A3}, as well as the ``BoNuS'' experiment
\cite{BoNuS12}, which minimizes nuclear corrections in semi-inclusive
DIS from deuterium by tagging slow, backward protons that effectively
guarantee scattering from a nearly-free neutron.
In addition, parity-violating DIS on a hydrogen target \cite{SOLID}
will yield a new combination of $u$ and $d$ PDFs at large $x$, free
of any nuclear corrections (see Ref.~\cite{CJ11} for further details).
In the meantime, however, it is important to establish the limitations
that the current uncertainties on the $d$ quark PDF at large $x$ place
on the calculation of observables which may be sensitive to these.
In the following we shall illustrate the impact of PDF uncertainties
at large $x$ arising specifically from the dependence on the model of
nuclear corrections in the deuteron.  The choice of CJ PDFs for this
purpose is merely for convenience, as these are the only PDFs available
that explicitly quantify the nuclear model dependence.

%%%%%%%%%%%%%%%%%%%%%%%%%%%%%%%%%%%%%%%%%%%%%%%%%%%%%%%%%%%%%%%%%%%%%%%%%
\section{$W$ and $Z$ boson production}
\label{sec:WZ}

In this section we discuss the effects of PDF uncertainties at large $x$
on the $W$ and $Z$ boson cross sections, and possible constraints on
these obtained from measurements at large rapidities.  Earlier studies
probing the sensitivity of weak boson production to PDF uncertainties
were explored in Refs.~\cite{Berger89, MST94, CTEQ3, Peng97, Alekhin10,
CT10, MSTW09, NNPDF11, NNPDF12}.
The discussion here is not meant to provide an exhaustive account of
detailed aspects of $W$ boson production, but simply highlight the
fact that nuclear corrections in deuterium are an important source of
PDF uncertainty at large $x$ that has not been addressed in earlier
analyses.
To begin with we shall review the general formulas for the inclusive
weak boson production cross sections in hadronic collisions relevant
to current collider experiments.

% .......................................................................
\subsection{Cross sections}
\label{sec:xsec}

Hadron--hadron collisions involve at least two interacting partons,
one from the hadron ``beam'' and one from the ``target'', with momentum
fractions $x_1$ and $x_2$, respectively.  At fixed center of mass energy
$\sqrt{s}$ and boson rapidity
\begin{equation}
y = {1 \over 2} \ln \left( {E + p_z \over E - p_z} \right),
\end{equation}
where $E$ and $p_z$ are the boson energy and longitudinal momentum
in the hadron center of mass frame, the parton momentum fractions
are given (at leading order in the strong coupling constant) by
\begin{equation}
x_{1, 2} = \frac{M}{\sqrt{s}}\, e^{\pm y},
\label{eq:x1x2}
\end{equation}
where $M$ is the mass of the produced boson.  The absolute value of the
rapidity thus ranges from 0 up to $|y|_{\rm max} = \log(\sqrt{s}/M)$.
For inclusive $W^+$ production in $pp$ or $p\bar p$ collisions,
for example, the cross sections (to leading order and neglecting
heavy quarks) are given by \cite{ColPhy}
\begin{subequations}
\label{eq:xsecW}
\begin{eqnarray}
\label{eq:xsecWLHC}
\frac{d\sigma}{dy}\left( pp \to W^+X \right)
&=& \frac{2\pi G_F}{3\sqrt{2}} x_1 x_2
\left( \cos^2{\theta_C}
  \left[ u(x_1)\overline{d}(x_2) + \overline{d}(x_1)u(x_2) \right]
\right. 						\nonumber\\
& & \hspace*{1.8cm} +
\left. \sin^2{\theta_C}
  \left[ u(x_1)\overline{s}(x_2) + \overline{s}(x_1)u(x_2) \right]
\right),						\\
\frac{d\sigma}{dy}\left( p\overline{p} \to W^+X \right)
&=& \frac{2\pi G_F}{3\sqrt{2}} x_1 x_2
\left( \cos^2{\theta_C}
  \left[ u(x_1)d(x_2) + \overline{d}(x_1)\overline{u}(x_2) \right]
\right. 						\nonumber\\
& & \hspace*{1.8cm} +
\left. \sin^2{\theta_C}
  \left[ u(x_1)s(x_2) + \overline{s}(x_1)\overline{u}(x_2) \right]
\right),
\end{eqnarray}
\end{subequations}
where $G_F$ is the Fermi constant, and $\theta_C$ is the Cabibbo
mixing angle.
The $W^-$ differential cross sections are similar to those in
Eqs.~(\ref{eq:xsecW}), but with quark PDFs replaced by the
corresponding antiquark PDFs.  Consequently the $W^\pm$ cross
sections in $p\bar p$ collisions are related by
$(d\sigma_{W^+}/dy)(y) = (d\sigma_{W^-}/dy)(-y)$,
and hence are equivalent when integrated over rapidity.
For $pp$ collisions the individual $W^+$ and $W^-$ cross sections
are symmetric with respect to $y \to -y$, but otherwise unrelated.

Similarly, the leading order, light quark cross sections for $Z$ boson
production in $pp$ or $p\bar p$ collisions are given by \cite{ColPhy}
\begin{subequations}
\label{eq:xsecZ}
\begin{eqnarray}
\frac{d\sigma}{dy}\left( pp \to ZX \right)
&=&\frac{2\pi G_F}{3\sqrt{2}} \sum_q
\left[ \left( g_V^q \right)^2 + \left( g_A^q \right)^2 \right]
x_1\, x_2\,
\big( q(x_1) \overline{q}(x_2) + \overline{q}(x_1) q(x_2) \big),  \\
\frac{d\sigma}{dy}\left( p\overline{p} \to ZX \right)
&=&\frac{2\pi G_F}{3\sqrt{2}} \sum_q 
\left[ \left( g_V^q \right)^2 + \left( g_A^q \right)^2 \right]
x_1\, x_2\,
\big( q(x_1) q(x_2) + \overline{q}(x_1) \overline{q}(x_2) \big),
\end{eqnarray}
\end{subequations}
where $g_V^q = t_3^q - 2 e_q \sW$ and $g_A^q = t_3^q$
are the vector and axial-vector couplings of the $Z$ boson to quark $q$
\cite{PDG10}, with $e_q$ and $t_3^q$ the electromagnetic charge and weak
isospin of the quark, respectively, and $\theta_W$ the weak mixing angle.
The symmetry properties of the differential $Z$ cross sections are such
that $(d\sigma_Z/dy)(y) = (d\sigma_Z/dy)(-y)$ for both $pp$ and $p\bar p$
collisions.
Note that in the conventions of Ref.~\cite{ColPhy} the couplings
$g_{V,A}^q$ are two times smaller than the standard ones in
Ref.~\cite{PDG10}.  In the convention used here the couplings
$(g_V^q)^2+(g_A^q)^2$ in Eqs.~(\ref{eq:xsecZ}) are equal to
$5/18 + \Delta(1+\Delta)/9$ and $13/36 + \Delta(1+\Delta/4)/9$
for $u$ and $d$ quarks, respectively, where $\Delta = 1 - 4\sW$.
Since $\Delta \approx 0$, the effective strengths of the $Z$ boson
couplings to $u$ and $d$ quarks are therefore similar.

While the expressions in Eqs.~(\ref{eq:xsecW}) and (\ref{eq:xsecZ})
are given at leading order, in practice we compute all cross sections
at NLO, including heavy quarks.  The leading order expressions give
the dominant contributions, however, and are instructive in clearly
illustrating that in $pp$ collisions, for instance, the $W^+$ cross
section at large rapidity is mostly dependent on the $u$ quark, while
the $W^-$ cross section depends mostly on the $d$.  Since the nuclear
corrections discussed in Sec.~\ref{sec:du} induce the greatest
uncertainty into the large-$x$ $d$ quark PDF, one can immediately
deduce that $W^-$ production in $pp$ scattering will be most affected
by these uncertainties, while $W^+$ production will be relatively inert.
For $p\bar p$ collisions, the large-$x$ PDF uncertainties will affect
$W^-$ cross sections at large positive rapidities, or equivalently
$W^+$ cross sections at negative rapidities.

In the following sections we will compute the $W$ and $Z$ boson cross 
sections numerically to study their sensitivity to PDF uncertainties
at large $x$.  All calculations will be for $pp$ collisions at the
LHC with $\sqrt{s} = 7$~TeV, and for $p\overline{p}$ collisions at the
Tevatron with $\sqrt{s} = 1.96$~TeV.

% .......................................................................
\subsection{$Z$ bosons}

\begin{figure}[ht]
% \rotatebox{-90}{\includegraphics[width=6cm]{plots/LHC.rap.Z.ratio.eps}}\hspace*{-0.5cm}%
% \rotatebox{-90}{\includegraphics[width=6cm]{plots/Tev.rap.Z.ratio.eps}}
\rotatebox{-90}{\includegraphics[width=6cm]{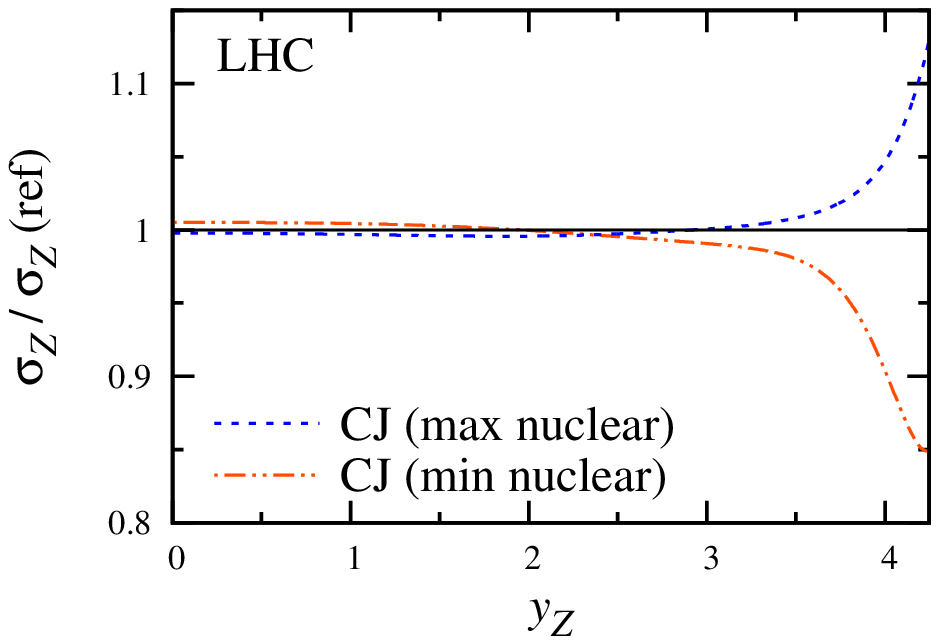}}\hspace*{-0.5cm}%
\rotatebox{-90}{\includegraphics[width=6cm]{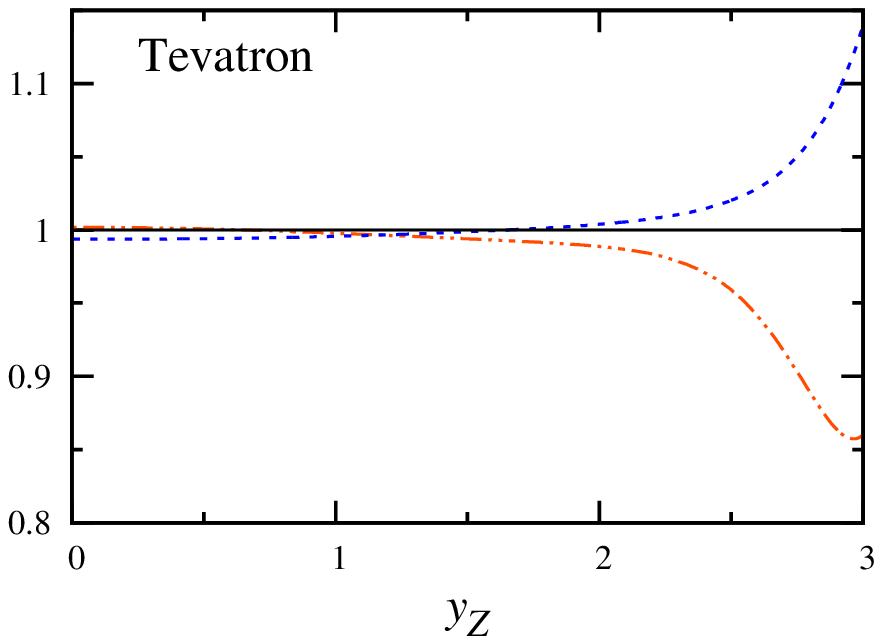}}
\caption{Differential $Z$ boson cross section as a function of the $Z$
	rapidity $y_Z$, computed from CJ PDFs with maximum (blue dashed)
	and minimum (red dot-dashed) nuclear corrections, relative to the
	reference cross section $\sigma_Z(\rm ref)$ calculated using the
	central CJ PDF set \cite{CJ11}.
	The cross sections are computed for $pp$ collisions at the LHC
	with $\sqrt{s}=7$~TeV {\bf (left)} and for $p\overline{p}$
	collisions at the Tevatron with $\sqrt{s}=1.96$~TeV {\bf (right)}.}
\label{fig:yZ}
\end{figure}

The sensitivity of the differential $Z$ boson cross section to the
different PDF behaviors at large $x$ is illustrated in Fig.~\ref{fig:yZ}
as a function of the $Z$ boson rapidity $y_Z$, for LHC and Tevatron
kinematics.  The cross sections are computed from the CJ PDFs \cite{CJ11}
with minimal and maximal nuclear corrections, relative to a reference 
cross section computed from the central PDFs as described in
Sec.~\ref{sec:du}.

The behavior of the cross section ratios is qualitatively similar at both
the LHC and the Tevatron, with the main difference being the range of
rapidities accessible from the respective available energies $\sqrt{s}$.
At low rapidities the cross sections are relatively insensitive to
uncertainties in the large-$x$ PDFs, with differences of $\lesssim 1\%$
for $y_Z \lesssim 3$ at the LHC and $y_Z \lesssim 2$ at the Tevatron.
At larger rapidities, however, there is far greater sensitivity to
the large-$x$ behavior, particularly of the $d$ quark, leading to
$\approx 15\%$ uncertainty in the differential cross section for
$y_Z = 4$ at the LHC, and for $y_Z = 2.8$ at the Tevatron, which
correspond to parton fractions of $x \approx 0.7$.
As one approaches the kinematical thresholds of
% $y_{Z, \rm max} \approx 4.34$ at the LHC and
% $y_{Z, \rm max} \approx 3.07$ at the Tevatron, these uncertainties
$y_{Z, \rm max} \approx 4.3$ at the LHC and
$y_{Z, \rm max} \approx 3.1$ at the Tevatron, these uncertainties
increase dramatically, as would be expected from the $x \to 1$
behavior of the $d/u$ ratio in Fig.~\ref{fig:du}.

% .......................................................................
\subsection{$W$ cross sections and asymmetries}

\begin{figure}[hbt]
% \rotatebox{-90}{\includegraphics[width=6cm]{plots/LHC.rap.W+.ratio.eps}}\hspace*{-0.5cm}%
% \rotatebox{-90}{\includegraphics[width=6cm]{plots/Tev.rap.W+.ratio.eps}}
% \rotatebox{-90}{\includegraphics[width=6cm]{plots/LHC.rap.W-.ratio.eps}}\hspace*{-0.5cm}%
% \rotatebox{-90}{\includegraphics[width=6cm]{plots/Tev.rap.W-.ratio.eps}}
\rotatebox{-90}{\includegraphics[width=6cm]{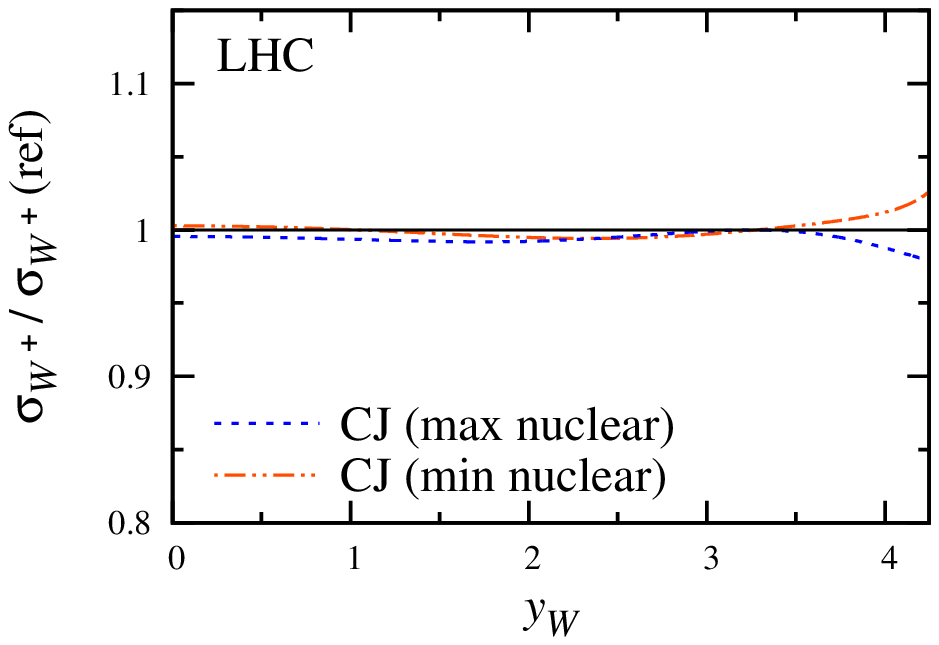}}\hspace*{-0.5cm}%
\rotatebox{-90}{\includegraphics[width=6cm]{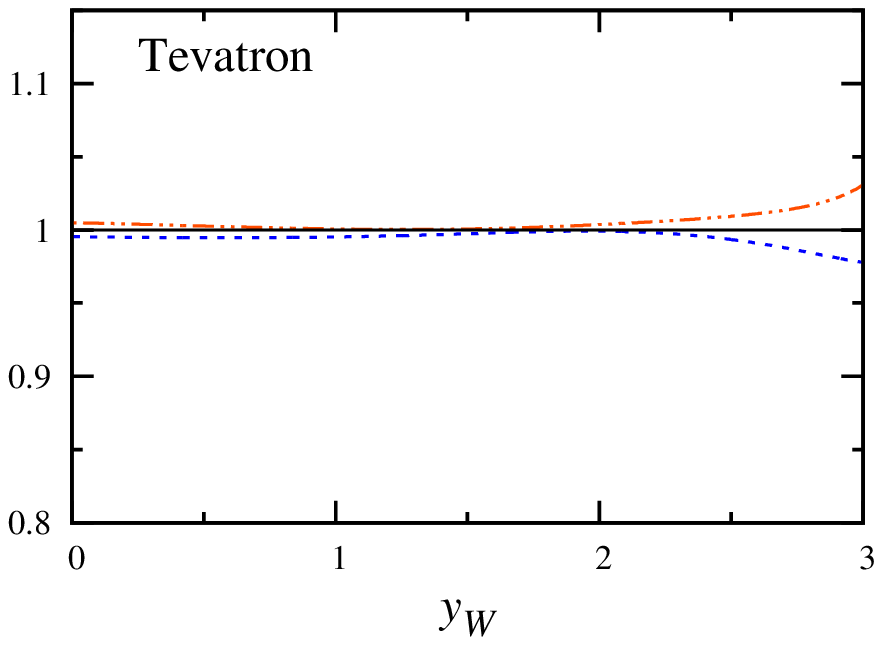}}
\rotatebox{-90}{\includegraphics[width=6cm]{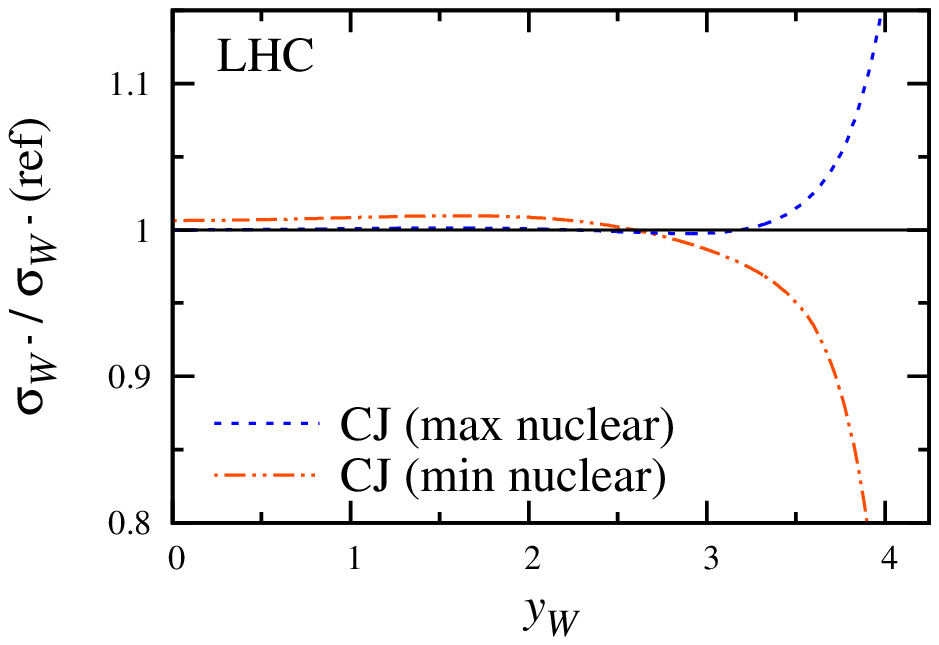}}\hspace*{-0.5cm}%
\rotatebox{-90}{\includegraphics[width=6cm]{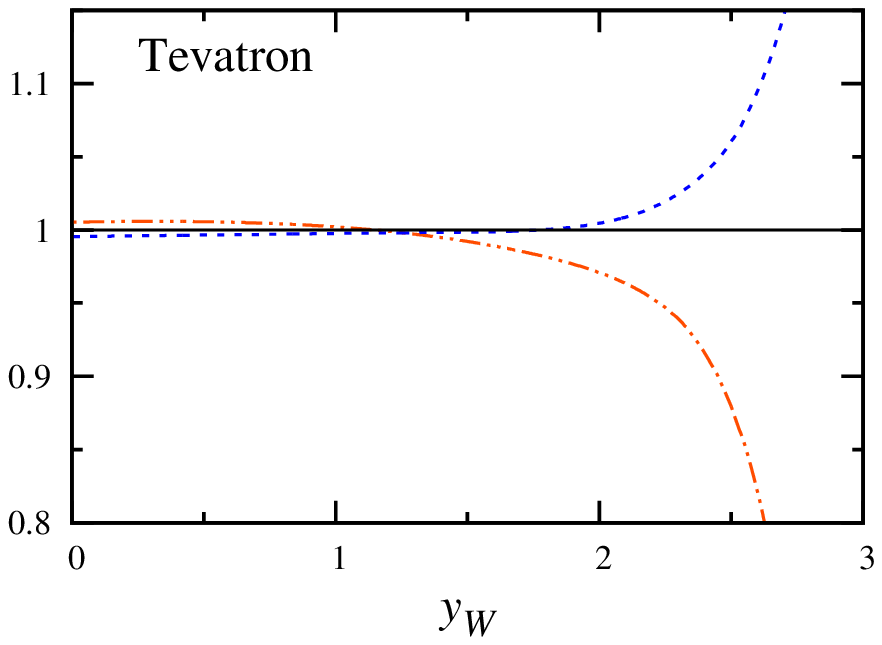}}
\caption{Differential $W^+$ {\bf (top)} and $W^-$ {\bf (bottom)} boson
	cross sections as a function of the $W$ rapidity $y_W$,
	computed from CJ PDFs with maximum (blue dashed) and minimum
	(red dot-dashed) nuclear corrections, relative to the reference
	cross sections $\sigma_{W^\pm}{\rm (ref)}$ calculated using the
	central CJ PDF set \cite{CJ11}.
        The cross sections are computed for $pp$ collisions at the
        LHC with $\sqrt{s}=7$~TeV {\bf (left)} and for $p\overline{p}$
        collisions at the Tevatron with $\sqrt{s}=1.96$~TeV {\bf (right)}.}
\label{fig:yW}
\end{figure}

The behavior of the $W$ boson differential cross sections as a function
of the $W$ rapidity $y_W$ is qualitatively similar to those for the $Z$,
but with some important differences, as Fig.~\ref{fig:yW} illustrates
for $W^+$ and $W^-$ production at LHC and Tevatron kinematics.  Again,
there is very little dependence on the large-$x$ PDF uncertainties at
low rapidity, but increasing sensitivity as the rapidity approaches its
% kinematic upper limit of $y_{W, \rm max} \approx 4.47$ at the LHC and
% $y_{Z, \rm max} \approx 3.19$ at the Tevatron.
kinematic upper limit of $y_{W, \rm max} \approx 4.5$ at the LHC and
$y_{Z, \rm max} \approx 3.2$ at the Tevatron.

For $W^+$ bosons, the cross section maintains relatively little
dependence on the large-$x$ nuclear corrections over the entire
rapidity range, barely reaching 4\% difference at $y_W \approx 4$
at the LHC, or $y_W \approx 3$ at the Tevatron.  In contrast, the
$W^-$ cross section shows an even stronger dependence on nuclear
corrections than the $Z$ cross section in Fig.~\ref{fig:yZ},
deviating significantly from unity for $y_W \gtrsim 3$ for the LHC
and $y_W \gtrsim 2$ for the Tevatron, and reaching upwards of 40\%
deviation at $y_W \approx 4$ and 3 for LHC and Tevatron kinematics,
respectively.
The greater sensitivity of the $W^-$ production cross section compared
with the $W^+$ can be understood from the dominance of the latter
by the $u$ quark PDF at large $x$, which is relatively insensitive
to the nuclear correction uncertainties.
The enhancement of the $W^-$ cross section at large $y_W$ for the CJ
PDFs with maximum nuclear corrections, and corresponding suppression
of the CJ PDFs with minimum nuclear corrections, relative to the
central CJ fits essentially follows the trend of the $d$ quark PDF in
Fig.~\ref{fig:du}.  The slight enhancement of the $W^+$ cross section at
large $y_W$ for the CJ PDFs with minimum nuclear corrections reflects the
anticorrelation of the $u$ quark PDF with respect to the $d$ observed in
Ref.~\cite{CJ11}.

\begin{figure}[t]
% \rotatebox{-90}{\includegraphics[width=6cm]{plots/LHC.W_asym.eps}}\hspace*{-0.5cm}%
% \rotatebox{-90}{\includegraphics[width=6cm]{plots/Tev.W_asym.eps}}
\rotatebox{-90}{\includegraphics[width=6cm]{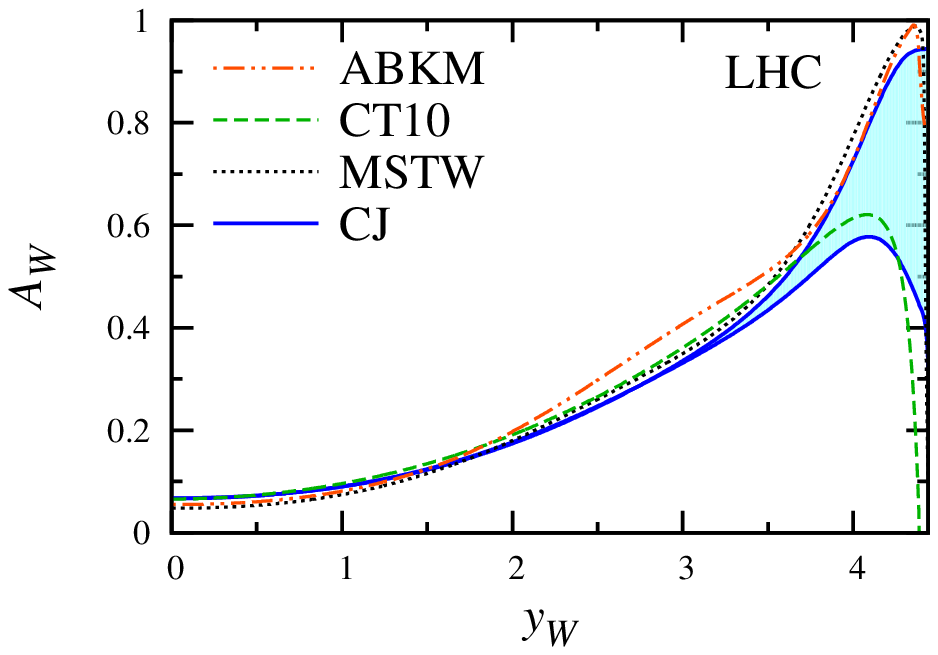}}\hspace*{-0.5cm}%
\rotatebox{-90}{\includegraphics[width=6cm]{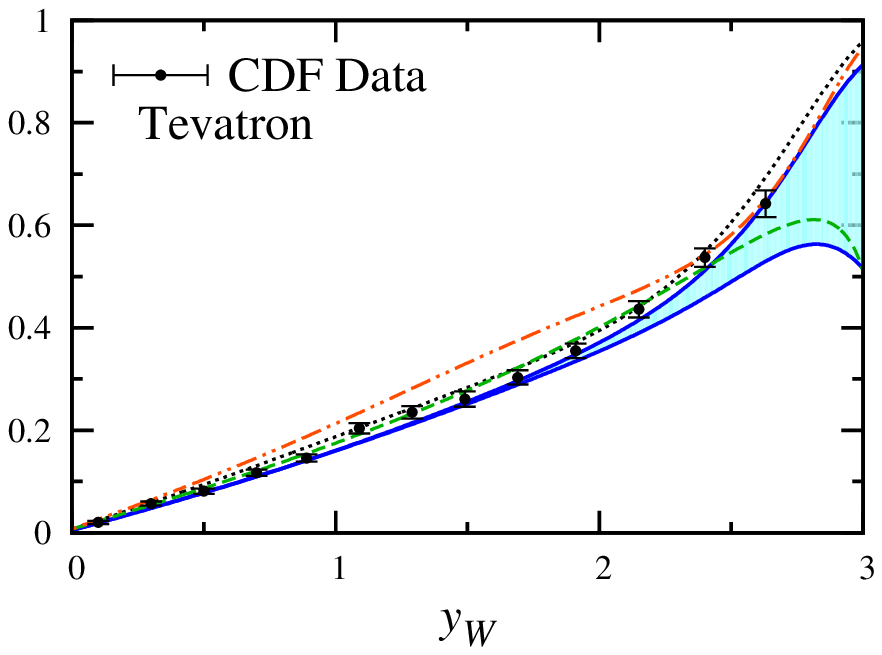}}
\caption{$W$ boson asymmetry $A_W$ as a function of the $W$ rapidity
	$y_W$ at LHC {\bf (left)} and Tevatron {\bf (right)} kinematics,
	computed from CJ PDFs \cite{CJ11} with minimum (upper blue solid)
	and maximum (lower blue solid) nuclear corrections.
	For comparison, the asymmetries using the ABKM \cite{Alekhin10}
	(red dot-dashed), CT10 \cite{CT10} (green dashed) and MSTW
	\cite{MSTW09} (black dotted) PDF sets are also shown.}
\label{fig:W_asym}
\end{figure}

Taking differences and sums of the $W^+$ and $W^-$ cross sections,
one can construct the $W$ boson asymmetry,
\begin{equation}
\label{eq:asy_W}
A_W = \frac{\sigma_{W^+}(y)-\sigma_{W^-}(y)}
	   {\sigma_{W^+}(y)+\sigma_{W^-}(y)},
\end{equation}
where $\sigma_{W^\pm}(y) \equiv d\sigma_{W^\pm}/dy$.
The asymmetry is shown in \fref{W_asym} versus the $W$ rapidity at
the LHC and Tevatron for the CJ PDFs with maximum and minimum nuclear
corrections.  A clear deviation between the two PDF fits becomes
visible at $y_W \gtrsim 3.5$ for the LHC and $y_W \gtrsim 2$ for
the Tevatron, corresponding to one of the partons carrying momentum
fractions $x \approx 0.4$ and $x \approx 0.35$, respectively.
Data on $W$ boson asymmetries may therefore provide constraints on
the $d/u$ quark distribution ratio already at these moderate values
of $x$.  In particular, comparison with the CDF data \cite{AW:CDF}
in Fig.~\ref{fig:W_asym} illustrates a preference for larger $A_W$
values at $y_W \gtrsim 2$, which corresponds to $x_1 \gg x_2$.
As observed in Ref.~\cite{CJ11}, since the asymmetry at large $y_W$
can be approximated by
\begin{equation}
\label{eq:asy_large_yW}
\hspace*{1cm}
A_W\ \approx\ \frac{d(x_2)/u(x_2) - d(x_1)/u(x_1)}
		   {d(x_2)/u(x_2) + d(x_1)/u(x_1)},
\hspace*{2cm} [x_1 \gg x_2],
\end{equation}
this would suggest a smaller $d/u$ ratio at large $x_1$, as would arise
for the CJ PDFs with minimum nuclear corrections.

For comparison, the asymmetries calculated from the ABKM
\cite{Alekhin10}, CT10 \cite{CT10} and MSTW \cite{MSTW09}
PDF sets are also shown in Fig.~\ref{fig:W_asym}.
Each parametrization shows good agreement with the Tevatron CDF data,
with the exception of the ABKM fit, which overestimates the asymmetries
at intermediate rapidities, $y_W \approx 1-2$.  This may be due to the
$W$ asymmetry data not being fitted directly in the ABKM analysis.
At large rapidity the spread in the various PDF sets is comparable to
the difference between the CJ PDFs with minimal and maximal nuclear
corrections.  However, we stress that the origin of the differences
between the PDF sets is unrelated to the difference between the two
CJ PDF sets.  Had the various non-CJ PDFs sets included nuclear
uncertainties in their analysis, each one would have a corresponding
nuclear uncertainty band similar to the one in Fig.~\ref{fig:W_asym},
and the combined spread between them would subsequently be
significantly larger.

% .......................................................................
\subsection{Lepton Asymmetries}
\label{sec:lep_asy}

Experimentally, measurement of $W$ bosons asymmetries requires
reconstruction of the $W$ boson distributions from their leptonic decays,
$W^+ \to l^+ \nu_l$ and $W^- \to l^- \bar\nu_l$, with $l = e$ or $\mu$.
On the other hand, lepton charge asymmetries can be constructed
directly from the $W^\pm$ decay products and studied as a function
of the lepton pseudorapidity $\eta$, defined as
\begin{equation}
\eta = {1 \over 2}
	\ln \left( {|{\bm k}| + k_z \over |{\bm k}| - k_z} \right)
     = -\ln \tan{\theta \over 2},
\end{equation}
where ${\bm k}$ is the charged lepton momentum, and $\theta$ is the
angle between the lepton momentum and the beam axis in the center of
mass frame.  The lepton asymmetry is then given by 
\begin{equation}
\label{eq:lep_asy}
A_\eta = \frac{ \sigma_{l^+}(\eta) - \sigma_{l^-}(\eta) }
	      { \sigma_{l^+}(\eta) + \sigma_{l^-}(\eta) },
\end{equation}
where $\sigma_{l^\pm}(\eta) \equiv d\sigma_{l^\pm}/d\eta$ is the
differential cross section for the production and leptonic decay
of the $W^\pm$.

\begin{figure}[b]
% \rotatebox{-90}{\includegraphics[width=6cm]{plots/lept_asym3.eps}}\hspace*{-0.5cm}%
% \rotatebox{-90}{\includegraphics[width=6cm]{plots/lept_asym3_D0.eps}}
\rotatebox{-90}{\includegraphics[width=6cm]{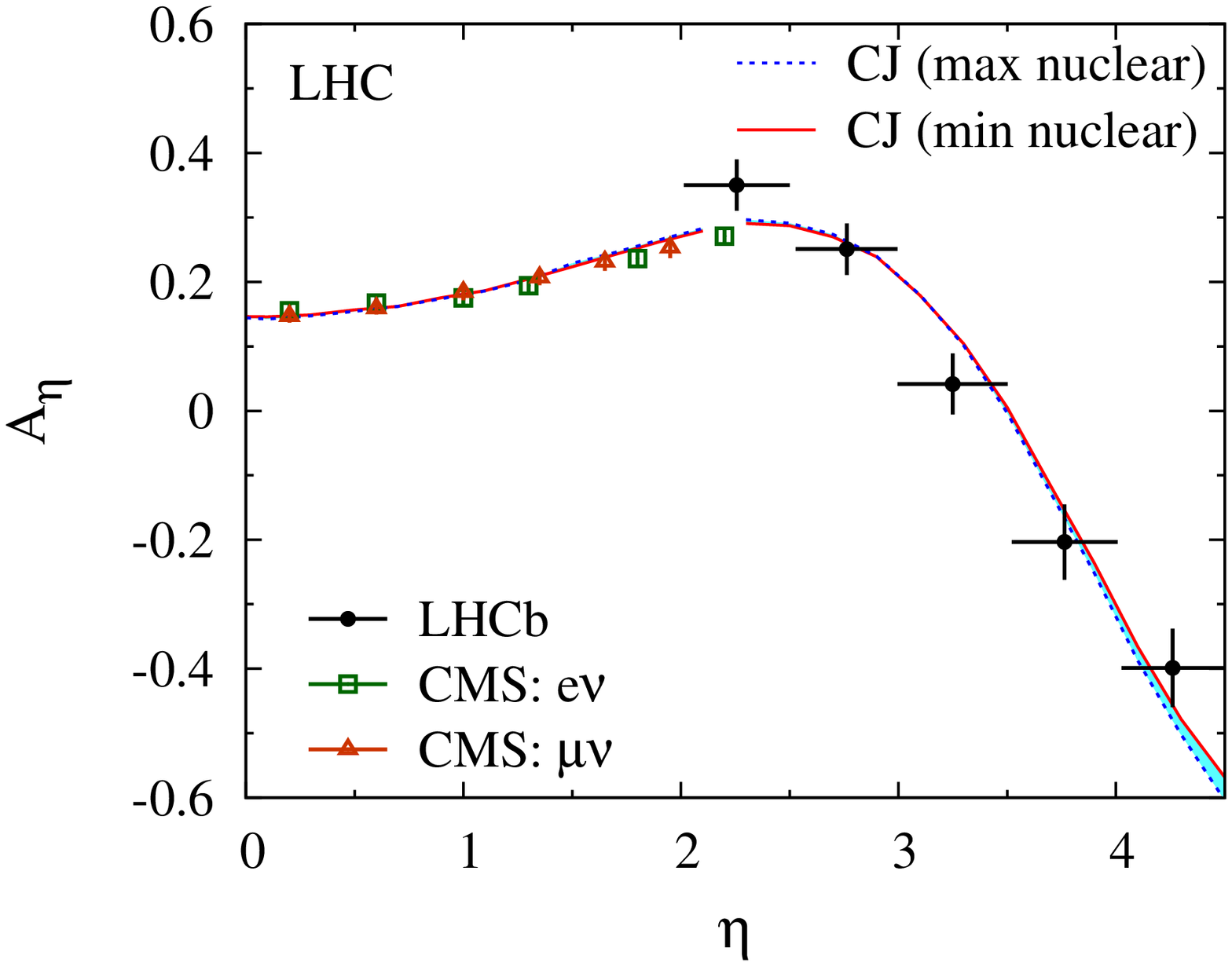}}\hspace*{-0.5cm}%
\rotatebox{-90}{\includegraphics[width=6cm]{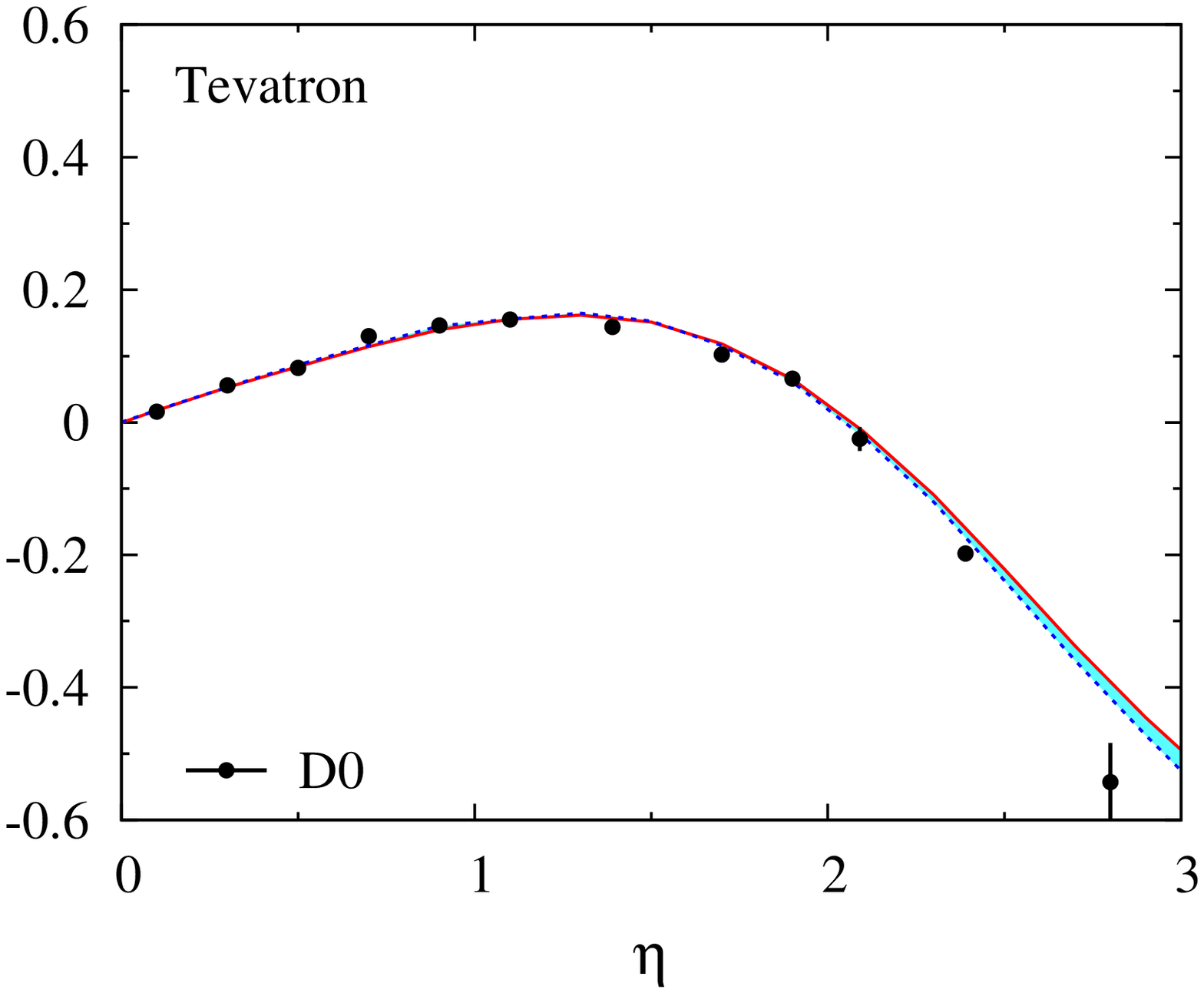}}
\caption{Lepton charge asymmetry $A_\eta$ as a function of lepton
        pseudorapidity $\eta$ at the LHC {\bf (left)} and Tevatron
        {\bf (right)}, computed from CJ PDFs \cite{CJ11} with maximum
        (blue dotted) and minimum (red solid) nuclear corrections.
        The calculations are compared with CMS $W \to e\nu_e$
	(red squares) and $W \to \mu\nu_\mu$ (blue triangles)
	data for a lepton transverse momentum cut $p_T > 25$~GeV
	\cite{Ae:CMS}, and preliminary LHCb data (black circles)
	for $p_T > 20$~GeV \cite{Ae:LHCb}, as well as with D0 data
	from the Tevatron $p_T > 25$~GeV \cite{Ae:D0}.}
\label{fig:lep_asy}
\end{figure}

Lepton asymmetry data from the D0 Collaboration \cite{Ae:D0} at Fermilab
are shown in Fig.~\ref{fig:lep_asy} as a function of the pseudorapidity
up to $\eta \approx 2.5$.  The data are integrated over lepton transverse
momenta $p_T > 25$~GeV, and compared with asymmetries computed from the
CJ PDFs with maximum and minimum nuclear corrections using the MCFM
(Monte Carlo for FeMtobarn processes) program \cite{MCFM}.
Good agreement is obtained between the calculated asymmetry and data,
although little sensitivity is evident to the large-$x$ nuclear
uncertainty in the PDFs observed in Fig.~\ref{fig:W_asym} until
$\eta \approx 3$.

Similar behavior is found for the new lepton charge asymmetry data
from the LHC.  Here lepton asymmetries have been measured by the
ATLAS \cite{Ae:ATLAS} and CMS \cite{Ae:CMS} Collaborations for
pseudorapidities $|\eta| \lesssim 2$, and preliminary results
from the LHCb Collaboration \cite{Ae:LHCb} extend the coverage to
$2 \lesssim \eta \lesssim 4.5$ \cite{Ae:LHC}.  The agreement of the
CJ PDFs with the LHC data is good over the entire range of $\eta$,
as Fig.~\ref{fig:lep_asy} illustrates for CMS $W \to e\nu$ and
$W \to \mu\nu$ data integrated over lepton transverse momenta
$p_T > 25$~GeV, and LHCb data with $p_T > 20$~GeV.  The dependence
on the large-$x$ behavior of PDFs, however, becomes visible only
for $\eta \gtrsim 4$.

The limited sensitivity of the lepton asymmetries to large-$x$ PDFs
is not surprising, given that the lepton asymmetry is computed by
convoluting the $W$ boson cross sections with the $W$ boson decay
distributions, which dilutes the sensitivity to regions where the PDFs
are small.  Although the lepton asymmetry data are clearly valuable
for constraining global PDF fits in general, greater sensitivity to
the large-$x$ behavior of the $d/u$ ratio may be possible through
the reconstruction of the $W$ boson asymmetries themselves.
Note, however, that the reconstruction of $W$ boson asymmetries is
limited by theoretical uncertainties such as the modeling of the $p_T$
distributions and higher-order resummation corrections (which also
affect the lepton asymmetries), as well as the choice of PDFs used
to compute the event reweighting coefficients in the reconstruction.

%%%%%%%%%%%%%%%%%%%%%%%%%%%%%%%%%%%%%%%%%%%%%%%%%%%%%%%%%%%%%%%%%%%%%%%%%
\section{Heavy $W'$ and $Z'$ bosons}
\label{sec:prime}

The production rate of any new heavy boson beyond the Standard Model
will naturally depend on its internal properties such as the spin.
Many possibilities have been canvassed for how such heavy bosons
can arise \cite{Rev:Rizzo06}, including as scalar excitations in
$R$-parity violating supersymmetry \cite{Z':spin0}, spin-1 Kaluza-Klein
excitations of Standard Model gauge bosons in the presence of extra
dimensions \cite{Z':spin1}, or as spin-2 excitations of the graviton
\cite{Z':spin2}.  On the other hand, if the new bosons are associated
with extensions of the Standard Model gauge group, their interactions
with fermions will resemble those of the $W$ and $Z$ bosons of the
electroweak theory, with different masses and couplings.

In this section we explore this latter possibility, and in particular
the sensitivity of the production cross sections to uncertainties in
PDFs at large $x$.  From Eq.~(\ref{eq:x1x2}) one can see that increasing
the mass will directly increase the relevant $x$ values, so that higher
mass bosons will more readily sample the high-$x$ region where the
nuclear uncertainties are more prominent.  In the calculations discussed
here we shall assume that the putative $W'$ and $Z'$ bosons have the
same properties as the Standard Model $W$ and $Z$ bosons, except for
their larger masses.  The cross sections will of course decrease rapidly
with increasing boson mass, so that the effects of the large-$x$ PDF
uncertainties will become more significant as the mass increases.
Of course the details of the predictions will change in more
sophisticated models in which the $W'$ and $Z'$ couplings are different
from those in the Standard Model; however, the simplified scenario
considered here is sufficient to illustrate the possible impact of
large-$x$ PDF uncertainties on new physics searches.

Currently the exclusion limit on the $W'$ mass from the Tevatron,
for couplings similar to Standard Model couplings, is
$M_{W'} > 1.12$~TeV at the 95\% confidence level with an
integrated luminosity of 5.3~fb$^{-1}$  \cite{W'lim_CDF}.
For the neutral $Z'$ boson the limits vary between
$M_{Z'} \gtrsim 800$~GeV and $M_{Z'} \gtrsim 1$~TeV \cite{Z'lim_CDF},
depending on the Standard Model extension considered \cite{PDG10}.
The latest results from the LHC place the limits for the $W'$ mass
at $M_{W'} > 2.15$~TeV \cite{W'lim_LHC} and for the $Z'$ mass at
$M_{Z'} > 1.83$~TeV \cite{Z'lim_LHC} in the Sequential Standard
Model, with the same couplings to fermions as for the $W$ and $Z$.
A similar constraint on the $Z'$ mass in grand unified theories is also
obtained from measurements of atomic parity violation in $^{133}$Cs
\cite{APV}.

% .......................................................................
\subsection{Rapidity distributions}

The differential cross sections for heavy $W'$ and $Z'$ bosons will
generally have similar behavior as a function of rapidity to those of
the physical boson cross sections in Figs.~\ref{fig:yZ} and \ref{fig:yW},
except for a smaller rapidity range, with the large-$x$ region emphasized
more strongly for increasing boson mass.
According to the study by Erler {\it et al.} \cite{Erler11}, $pp$
collisions at the LHC with $\sqrt{s}=7$~TeV will be sensitive to $Z'$
masses up to $M_{Z'} \approx 2.1 - 2.7$~TeV for luminosities between
30~fb$^{-1}$ and 300~fb$^{-1}$, corresponding to the so-called
``low-luminosity'' and ``high-luminosity'' LHC scenarios, respectively.
(With $\sqrt{s}=14$~TeV the mass limits would vary between
$M_{Z'} \approx 3.6$ and 4.6~TeV.)
For $p\bar p$ collisions at the Tevatron, an energy of $\sqrt{s}=2$~TeV
with a luminosity of 10~fb$^{-1}$ would be expected to allow sensitivity
to $Z'$ masses up to $\approx 1$~TeV \cite{Erler11}.
In this section we therefore consider $Z'$ (and $W'$) masses up to
1~TeV and 3~TeV for Tevatron and LHC kinematics, respectively.

\begin{figure}[ht]
% \rotatebox{-90}{\includegraphics[width=6cm]{plots/LHC.rap.Z_123TeV.ratio.eps}}\hspace*{-0.5cm}%
% \rotatebox{-90}{\includegraphics[width=6cm]{plots/Tev.rap.Z_123TeV.ratio.eps}}
% \rotatebox{-90}{\includegraphics[width=6cm]{plots/LHC.rap.W+_123TeV.ratio.eps}}\hspace*{-0.5cm}%
% \rotatebox{-90}{\includegraphics[width=6cm]{plots/Tev.rap.W+_123TeV.ratio.eps}}
% \rotatebox{-90}{\includegraphics[width=6cm]{plots/LHC.rap.W-_123TeV.ratio.eps}}\hspace*{-0.5cm}%
% \rotatebox{-90}{\includegraphics[width=6cm]{plots/Tev.rap.W-_123TeV.ratio.eps}}
\rotatebox{-90}{\includegraphics[width=6cm]{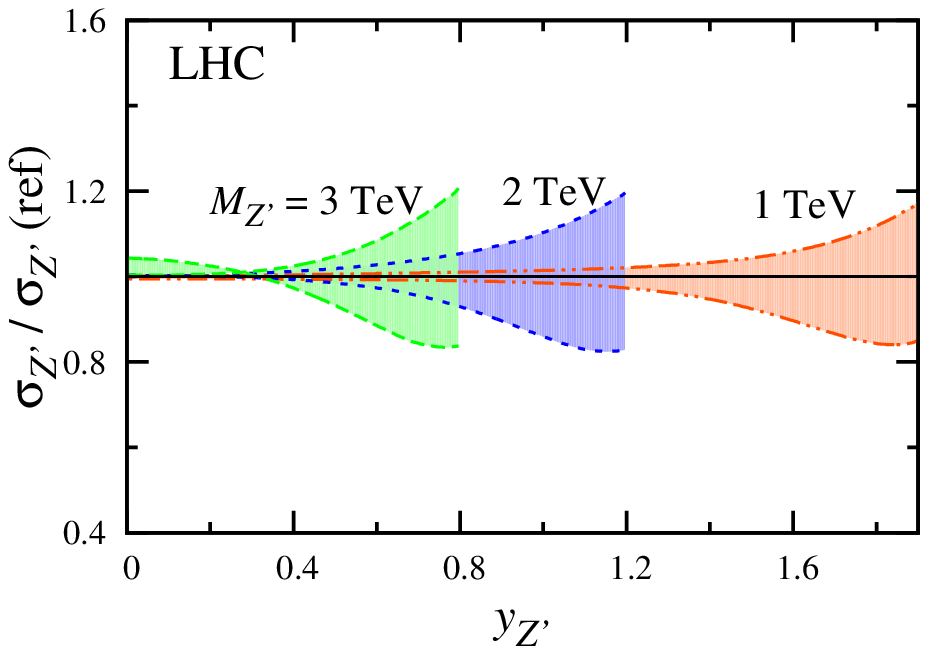}}\hspace*{-0.5cm}%
\rotatebox{-90}{\includegraphics[width=6cm]{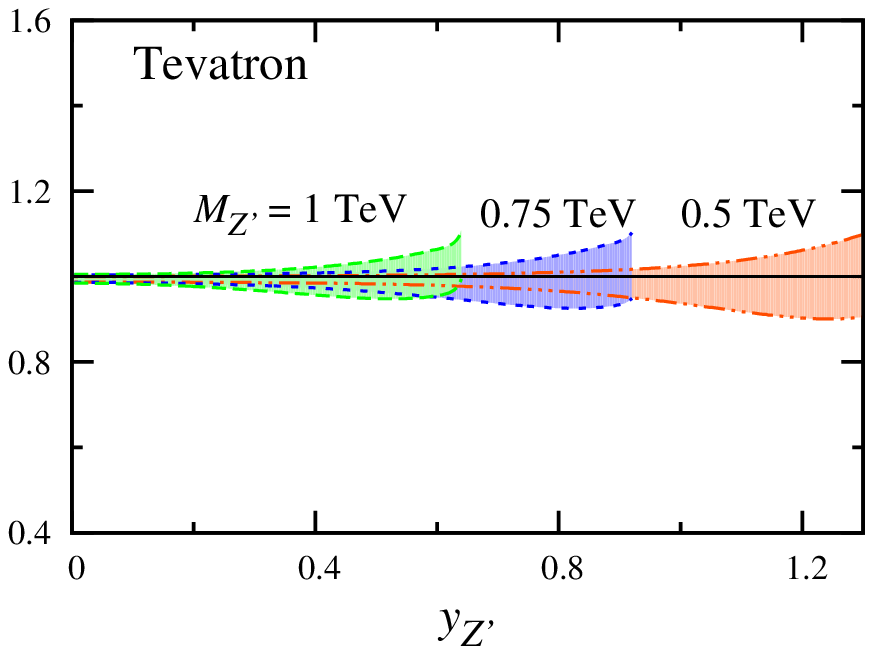}}
\rotatebox{-90}{\includegraphics[width=6cm]{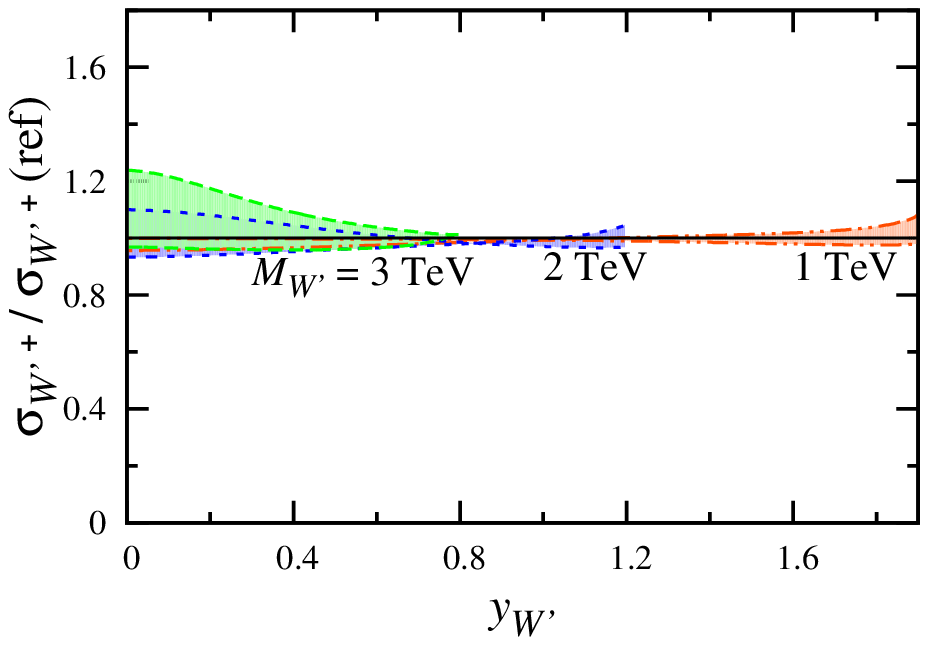}}\hspace*{-0.5cm}%
\rotatebox{-90}{\includegraphics[width=6cm]{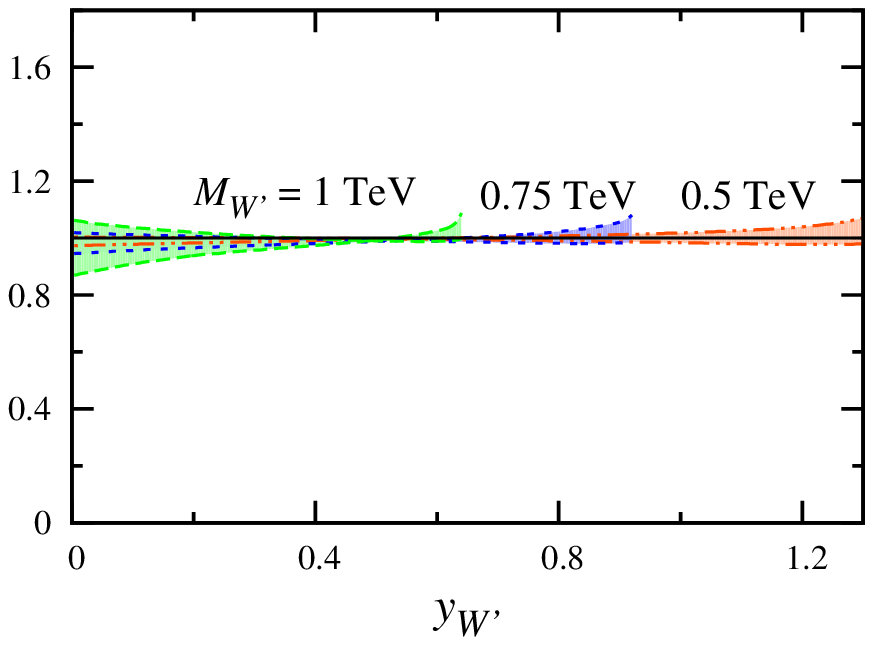}}
\rotatebox{-90}{\includegraphics[width=6cm]{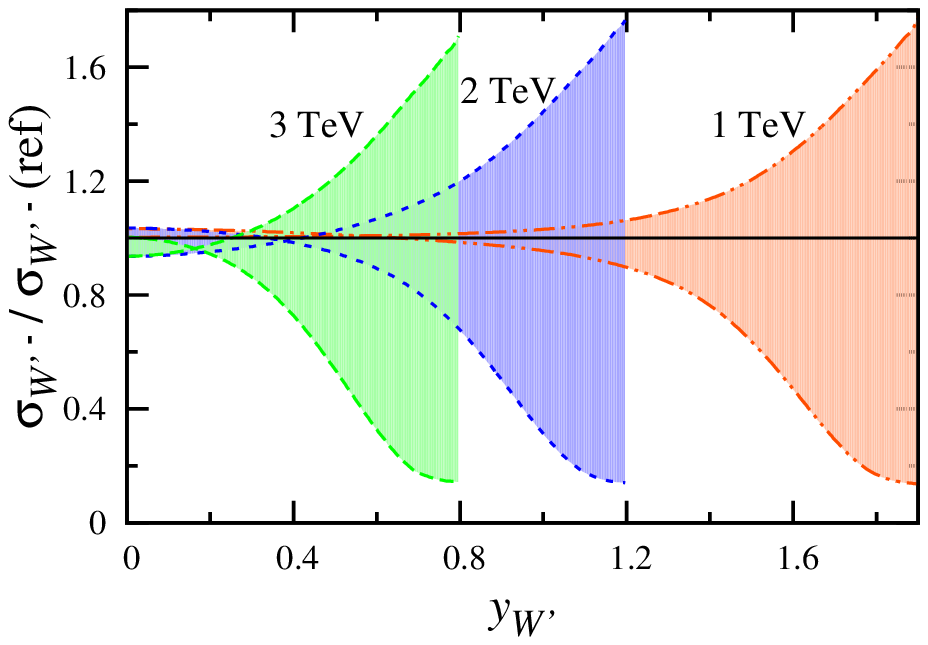}}\hspace*{-0.5cm}%
\rotatebox{-90}{\includegraphics[width=6cm]{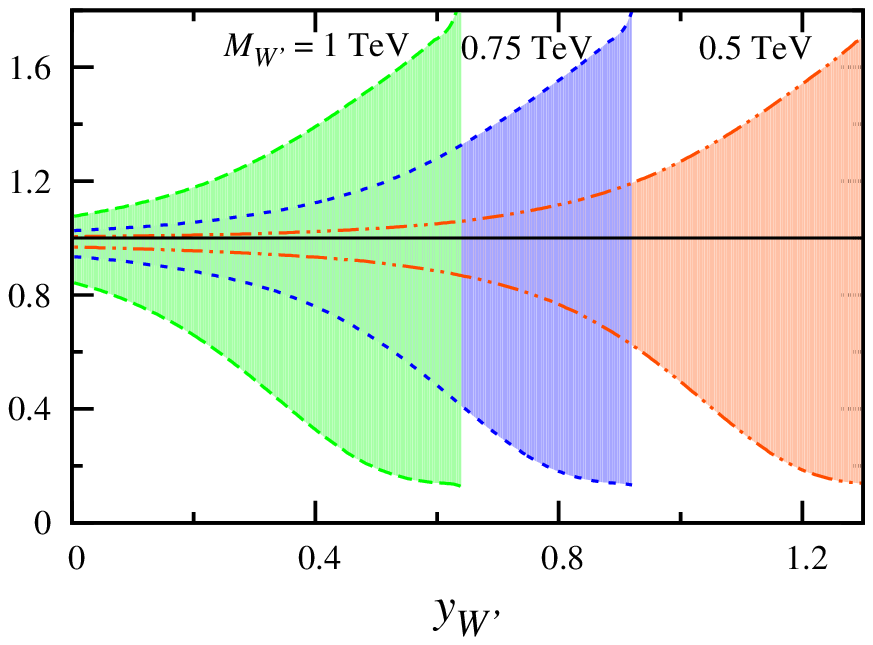}}
\caption{Differential $Z'$ {\bf (top)}, $W'^+$ {\bf (center)} and
	$W'^-$ {\bf (bottom)} cross sections as a function of the
	rapidity, computed from CJ PDFs with maximum and minimum
	nuclear corrections, relative to the reference cross sections
	$\sigma_{Z',W'} (\rm ref)$ calculated using the central CJ PDF
	set \cite{CJ11}.
	The LHC cross sections {\bf (left)} are computed for boson masses
	$M_{Z',W'} = 1$~TeV (red dot-dashed), 2~TeV (short-dashed)
	and 3~TeV (long-dashed) with $\sqrt{s}=7$~TeV,
	while the Tevatron ratios {\bf (right)} are shown for
	$M_{Z',W'} = 0.5$~TeV (red dot-dashed), 0.75~TeV (short-dashed)
	and 1~TeV (long-dashed) with $\sqrt{s}=1.96$~TeV.}
\label{fig:yWZ'}
\end{figure}

From Eq.~(\ref{eq:x1x2}), larger boson masses naturally restrict the
kinematically accessible range of rapidities, so that at the LHC,
for example, a 1~TeV (3~TeV) $Z'$ boson can be produced at a maximum
rapidity of
$|y_{Z'}|_{\rm max} = 2.0$ (0.9), compared with 
$|y_{Z}|_{\rm max}  = 4.3$ for the Standard Model $Z$ boson.
The kinematic reach and sensitivity to large-$x$ PDFs is illustrated in
Fig.~\ref{fig:yWZ'}~(top) for the differential $Z'$ cross section ratio
as a function of $y_{Z'}$, for masses $M_{Z'} = 1$, 2 and 3~TeV at the
LHC, and $M_{Z'} = 0.5$, 0.75 and 1~TeV at the Tevatron.
As the rapidities approach their kinematical thresholds for a given
$M_{Z'}$, the uncertainty in the differential cross sections increases
significantly, reaching about $30-40\%$ of the central CJ value for
$pp$ collisions at the LHC, and $15-20\%$ for $p\bar p$ collisions
at the Tevatron.

Because the couplings of the $Z'$ to $u$ and $d$ quarks are assumed
to be similar (see Sec.~\ref{sec:xsec}), the $Z'$ cross section at the
LHC depends on both the combinations
    $u(x_1) \overline u(x_2) + \overline u(x_1) u(x_2)$
and $d(x_1) \overline d(x_2) + \overline d(x_1) d(x_2)$.
However, since the $d/u$ and $\bar u/u$ ratios are $\ll 1$ at
large $x$ values, which are preferentially sampled for large $M_{Z'}$,
the contributions of $d$ quarks are suppressed relative to $u$ quarks.
Consequently the $Z'$ ratios in Fig.~\ref{fig:yWZ'} are only mildly
affected by uncertainties in quark PDFs at large $x$.
The $Z'$ production cross section in $p\overline p$ collisions at the
Tevatron, on the other hand, is determined predominantly by the product
$u(x_1) u(x_2)$, which is well constrained and independent of the
nuclear model for all $x_{1,2}$, and therefore has an even smaller
uncertainty.

For the $W'$ differential cross sections, the behavior as a function of
$y_{W'}$ is qualitatively different for $W'^+$ and $W'^-$ production,
shown in Fig.~\ref{fig:yWZ'}~(center) and (bottom), respectively, with
the latter displaying dramatically greater sensitivity to large-$x$ PDF
uncertainties.  This is clear from Eqs.~(\ref{eq:xsecW}), where for $pp$
collisions the dominant contribution to the $W'^+$ cross section depends
on the products $u(x_1) \overline d(x_2)$ and $\overline d(x_1) u(x_2)$.
While the $u$ quark PDF is insensitive to the nuclear corrections,
the $\overline d$ distribution varies considerably with the nuclear
model, especially at larger values of $x$.
At high rapidity the $\bar d/u$ ratio is small, and the cross section
is determined by the $u$ PDF with $x_1$ large and the $\overline d$ PDF
with $x_2$ small, both of which are well constrained.
For $y_{W'} \to 0$, on the other hand, one has
$x_1 = x_2 \approx 0.14$ for $M_{W'} = 1$~TeV and
$x_1 = x_2 \approx 0.42$ for $M_{W'} = 3$~TeV, at which the
$\overline d$ PDF has significantly greater uncertainty than the $u$,
yielding up to $\approx 25\%$ uncertainties in the cross section.
Similarly for $p\overline p$ collisions at the Tevatron, the $W'^+$
cross section is dominated by the combination $u(x_1) d(x_2)$, which
for $x_1 \gg x_2$ at high rapidity is relatively well constrained.
At central rapidity, with $x_1 = x_2 \approx 0.25~(0.5)$ for
$M_{W'} = 0.5~(1)$~TeV, the uncertainties in the cross section
remain within the $\approx 10\%$ level.

In contrast, the $W'^-$ cross section at high rapidity is determined by
the products $d(x_1) \overline u(x_2)$ and $\overline u(x_1) d(x_2)$
in $pp$ collisions, and $d(x_1) u(x_2)$ and $u(x_1) d(x_2)$
in $p \overline p$ collisions.
Consequently, uncertainties in the cross sections at large rapidity,
both at the LHC and the Tevatron, arise mainly from the $d$ quark
at large $x$, and exceed 100\% as the kinematic limit in $y_W$ is
approached.  Qualitatively, the growing uncertainty of the $W'^-$
cross section with increasing rapidity resembles the $W^-$ cross
section ratio at large $y_W$ in Fig.~\ref{fig:yW}.
At central rapidity, the uncertainty in the $pp$ cross section at the LHC
is of the order 10\%, arising mainly from the uncertainty in the $\bar u$
distribution.  At the Tevatron, the $p \overline p$ cross section is well
constrained for small boson mass, but for $M_{W'} > 0.75$~TeV, with
$x_1 = x_2 \gtrsim 0.3$, becomes increasingly sensitive to uncertainties
in the $d$ quark PDF, reaching about 20\% for $M_{W'} = 1$~TeV.

% .......................................................................
\subsection{Integrated cross sections}

Integrating over all rapidities, the resulting total $Z'$ cross section
computed from CJ PDFs with minimum and maximum nuclear corrections is
shown in \fref{msZ} as a function of the $Z'$ mass.  Relatively little
dependence on the PDFs is observed, with effects of $\lesssim 3\%$
observed for $M_{Z'} < 3$~TeV at the LHC and $M_{Z'} < 1$~TeV at the
Tevatron.  This is not surprising given that total cross sections are
dominated by contributions from low values of $y_{Z'}$, where the PDF
uncertainties are generally smaller than at high values of $y_{Z'}$,
at which the contributions are suppressed by the steeply falling PDFs
as $x \to 1$.  At larger $M_{Z'}$ values the uncertainties generally
increase, but are subject to greater fluctuations in the antiquark
distributions at high $x$, and hence are less reliable.

\begin{figure}[hb]
% \rotatebox{-90}{\includegraphics[width=6cm]{plots/LHC.ms.Z.ratio.eps}}\hspace*{-0.5cm}
% \rotatebox{-90}{\includegraphics[width=6cm]{plots/Tev.ms.Z.ratio.eps}}
\rotatebox{-90}{\includegraphics[width=6cm]{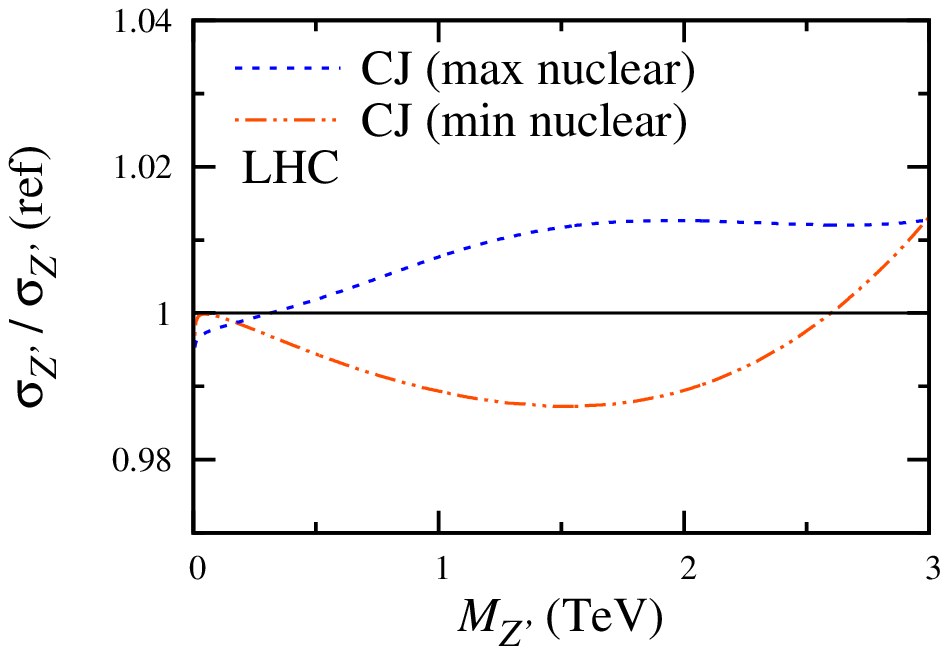}}\hspace*{-0.5cm}
\rotatebox{-90}{\includegraphics[width=6cm]{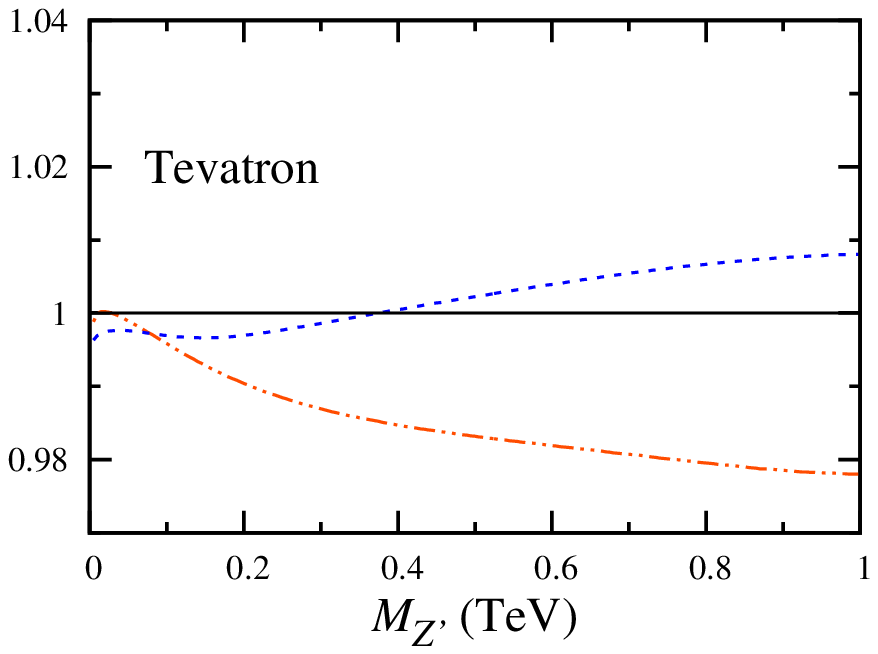}}
\caption{Integrated $Z'$ boson cross section from Fig.~\ref{fig:yWZ'} as
	a function of the $Z'$ mass, computed from CJ PDFs with minimum
	(red dot-dashed) and maximum (blue dashed) nuclear corrections,
	relative to the reference cross section $\sigma_{Z'}(\rm ref)$
	calculated using the central CJ PDF set \cite{CJ11},
	for LHC {\bf (left)} and Tevatron {\bf (right)} kinematics.}
\label{fig:msZ}
\end{figure}

\begin{figure}[t]
\parbox[t]{0.495\linewidth}{\flushleft
% \rotatebox{-90}{\includegraphics[width=5.9cm]{plots/LHC.ms.W+.ratio.eps}} \\
% \rotatebox{-90}{\includegraphics[width=5.9cm]{plots/LHC.ms.W-.ratio.eps}}
  \rotatebox{-90}{\includegraphics[width=5.9cm]{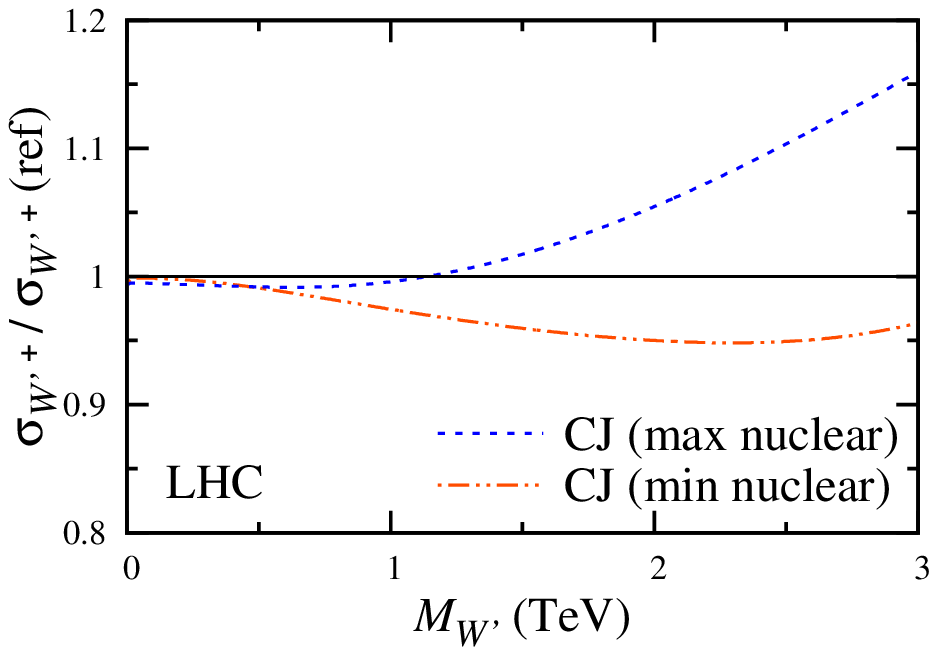}} \\
  \rotatebox{-90}{\includegraphics[width=5.9cm]{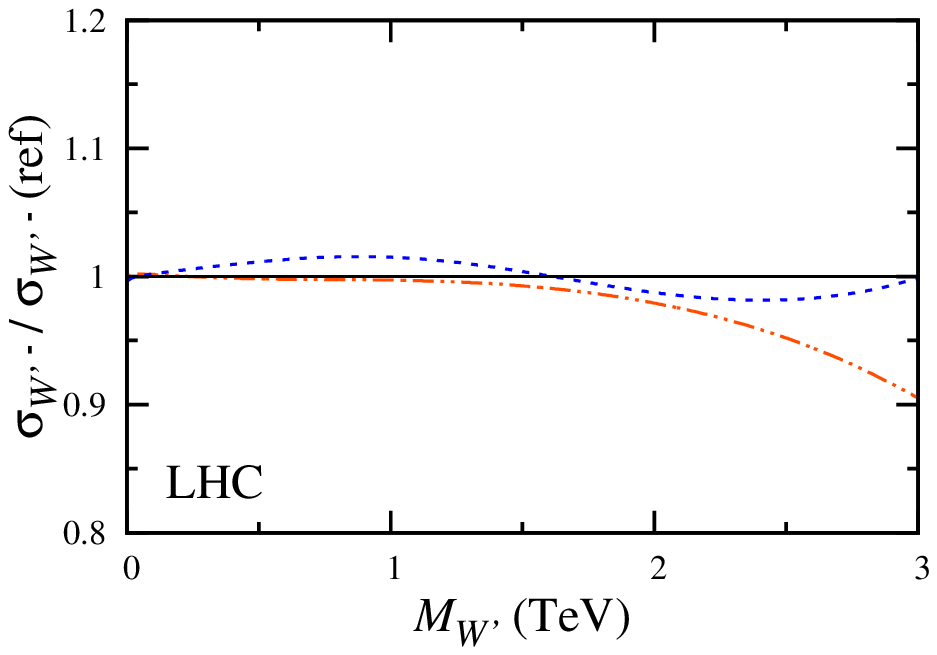}}
}
\parbox[t]{0.495\linewidth}{\flushright
% \rotatebox{-90}{\includegraphics[width=5.9cm]{plots/Tev.ms.W+.ratio.eps}}
  \rotatebox{-90}{\includegraphics[width=5.9cm]{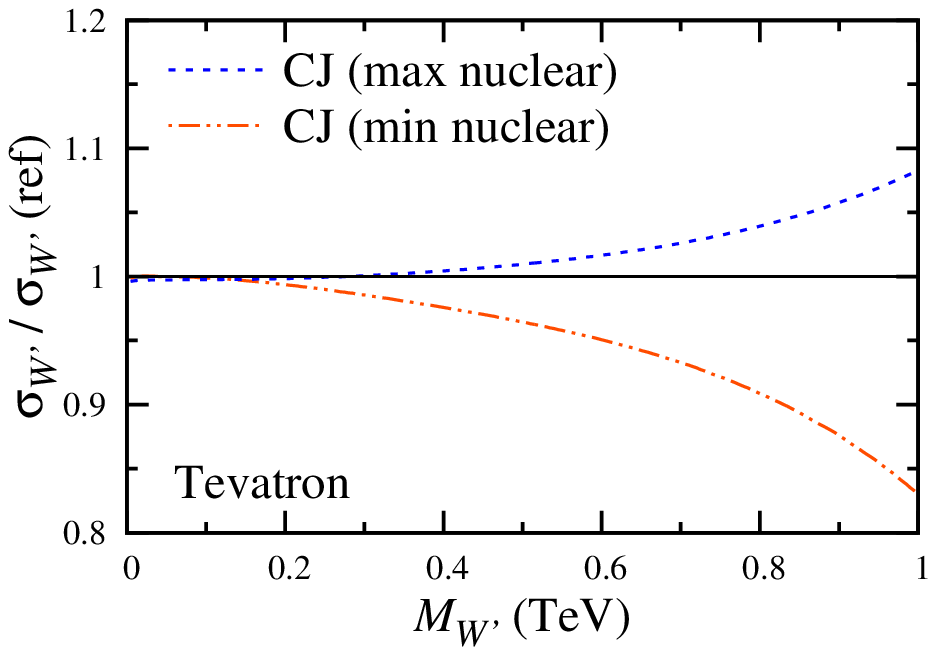}}
  \vskip 0.5cm \parbox[b]{0.9\linewidth}{
        \caption{As in Fig.~\ref{fig:msZ} but for $W'^+$ and $W'^-$ cross
	sections at the LHC {\bf (left)} and Tevatron {\bf (right)}.
	Note that the integrated cross sections for $W'^+$ and $W'^-$
	production in $p\overline p$ collisions at the Tevatron
	are identical.}
\label{fig:msW}
}
}
\end{figure}

The integrated cross sections for $W'$ bosons in Fig.~\ref{fig:msW} show
somewhat greater sensitivity to large-$x$ PDFs as a function of $M_{W'}$.
For $W'^+$ bosons produced in $pp$ collisions at the LHC the uncertainties
increase from $\lesssim 5\%$ for $M_{Z'} = 1$~TeV to $\approx 20\%$ for
$M_{Z'} = 3$~TeV.  This behavior stems directly from the increasing
uncertainty in the $\overline d$ antiquark PDF at large $x$ apparent in
the $W'^+$ rapidity distribution at low $y_{W'}$ in Fig.~\ref{fig:yWZ'}.
For $W'^-$ boson production in $pp$ scattering, the uncertainties in
the total cross section are smaller than for the $W'^+$ at low $M_{W'}$,
remaining $\lesssim 2\%$ for $M_{W'} < 2$~TeV, but increase to
$\approx 10\%$ at $M_{W'} = 3$~TeV due to the uncertainty in the
$\overline u$ quark.  The stronger dependence on the behavior of the
$d$ quark PDF at large $x$ apparent in the $W'^-$ differential cross
section at high rapidity in Fig.~\ref{fig:yWZ'} is mostly washed out
in the integrated cross section.

Because the $W'$ cross sections in $p\overline p$ collisions at
the Tevatron are determined by the products $u(x_1) d(x_2)$ and
$d(x_1) u(x_2)$ for $W'^+$ and $W'^-$, respectively, integrating
over rapidity samples all accessible values of $x_1$ and $x_2$,
so that the total $W'^+$ and $W'^-$ cross sections are equivalent.
The dependence of the integrated $W'$ cross sections on PDFs
essentially follows the $d$ quark distribution.  For masses
$M_{W'} \lesssim 0.5$~TeV there is little sensitivity to the
large-$x$ behavior of the PDFs, with $\lesssim 5\%$ uncertainty
in the cross section ratio, but increasing dependence at larger
$M_{W'}$, with $\approx 30\%$ uncertainty at $M_{W'} = 1$~TeV.

\begin{figure}[t]
% \rotatebox{-90}{\includegraphics[width=6cm]{plots/LHC.combined.eps}}
\rotatebox{-90}{\includegraphics[width=6cm]{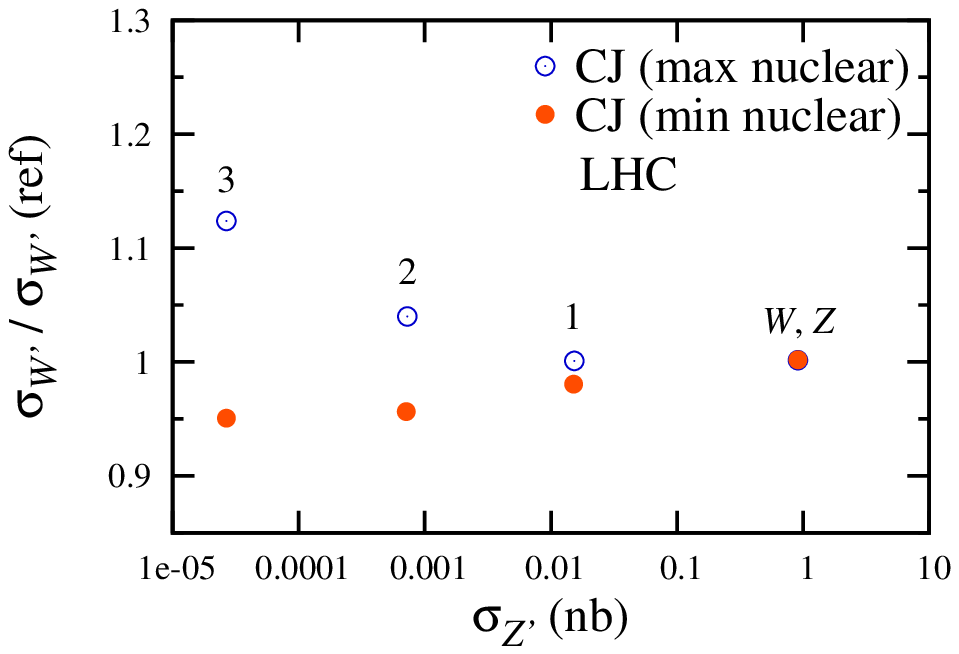}}
\caption{Ratio of $W'$ boson cross sections
	$\sigma_{W'}/\sigma_{W'}{\rm (ref)}$, where
	$\sigma_{W'} = \sigma_{W'^+} + \sigma_{W'^-}$,
	versus the $Z'$ boson cross section $\sigma_{Z'}$ for varying
	$Z'$ boson masses, indicated by the numbers (in TeV) next to
	the points (the points corresponding to the physical $W$ and
	$Z$ cross sections are labeled by ``$W, Z$'').
	The cross sections are computed from CJ PDFs with minimum
	(filled red circle) and maximum (open blue circle) nuclear
	corrections, relative to the reference cross section calculated
	from the central CJ PDFs \cite{CJ11}.}
\label{fig:comb}
\end{figure}

While the sensitivity of the $W'$ and $Z'$ cross sections to the
large-$x$ behavior of PDFs increases with increasing $W'$ and $Z'$
masses, the absolute values of the cross sections naturally fall with
increasing masses, some 3 orders of magnitude from 100~GeV to 3~TeV.
This is illustrated in \fref{comb}, where the ratio of the integrated
$W'^+ + W'^-$ cross sections computed from CJ PDFs with minimum and
maximum nuclear corrections, relative to the cross section with the
central CJ PDFs, is plotted versus the integrated $Z'$ cross section.
Here the ratio of the $W'$ to $Z'$ masses is kept constant in order
to study the effect of the increasing $W', Z'$ mass.  For larger boson
masses the impact of the large-$x$ PDF uncertainties clearly increases,
reflecting the trend observed in Figs.~\ref{fig:msZ} and \ref{fig:msW}.
Note that because the integrated $W'^+$ cross section is generally larger
than the $W'^-$ cross section (because of the larger $u$ distribution
compared with the $d$), the $\sigma_{W'}/\sigma_{W'}{\rm (ref)}$ ratio
in Fig.~\ref{fig:comb} generally follows the ratio of the $W'^+$ cross
sections in Fig.~\ref{fig:msW} for increasing boson mass.

%%%%%%%%%%%%%%%%%%%%%%%%%%%%%%%%%%%%%%%%%%%%%%%%%%%%%%%%%%%%%%%%%%%%%%%%%
\section{Conclusion}
\label{sec:conc}

In this paper we have explored the sensitivity of weak boson production
in hadronic collisions to parton distributions at large values of $x$.
At present there are large uncertainties in the $d$ quark distribution,
particularly above $x \approx 0.5$, arising from the model dependence of
nuclear corrections used when analyzing deuteron DIS data in global PDF
fits, which can impact cross section measurements at large rapidities.
The PDF uncertainties can also affect production cross sections of heavy
$W'$ and $Z'$ bosons beyond the Standard Model at central rapidities.

Using PDFs extracted from the recent CJ global fit \cite{CJ11}, we
find increasing sensitivity to the large-$x$ region for $Z$ boson
production in $p \overline p$ collisions at the Tevatron for $Z$
rapidities $y_Z \gtrsim 2$, and in $pp$ collisions at the LHC for
$y_Z \gtrsim 3$.
% but about a factor 2 smaller than current Tevatron data uncertainties,
Precision measurements of $Z$ boson cross sections at these rapidities,
at the Tevatron and particularly at the LHC with the LHCb experiment,
will be required to impact global fits of the $d$ quark and constrain
nuclear model uncertainties.

For charged weak bosons, the $W^+$ cross sections are mostly independent
of PDF uncertainties due to their preferential coupling to $u$ quarks,
whereas the $W^-$ cross sections display strong dependence on the $d$
quark uncertainties for $W$ rapidities $y_W \gtrsim 1.5$ at Tevatron and
$y_W \gtrsim 3$ at LHC kinematics.  Measurements of $W$ boson charge
asymmetries at large rapidities, such as those from the CDF Collaboration
at Fermilab \cite{AW:CDF}, thus provide strong constraints on the
behavior of the $d/u$ ratio at large $x$, although such measurements are
very challenging given the low rates expected in the relevant regions of
kinematics.
Direct reconstruction of $W$ boson asymmetries in the LHCb experiment
for $\sqrt{s}=7$~TeV would also be extremely valuable in providing
access to PDFs at $x \approx 1$.
We have also compared charged lepton asymmetries with data from D0
\cite{Ae:D0}, and from the CMS \cite{Ae:CMS} and LHCb Collaborations 
\cite{Ae:LHCb} at the LHC, finding good overall agreement with the
CJ PDFs, but weak sensitivity to large-$x$ PDFs.

The large-$x$ PDF uncertainties also affect the production rates
of heavy $W'$ and $Z'$ bosons, and the impact of these was studied
as a function of the boson mass for Standard Model couplings.
At high rapidity, the $W'^+$ cross section in $pp$ and $p\overline p$
collisions is mostly sensitive to the $u$ quark PDF at large $x$ and
therefore well constrained.  The $Z'$ cross section displays a mild
sensitivity to the $d$ quark PDF at large $x$, reaching upwards of 40\%
uncertainty at the kinematic rapidity limit.  The effects are even more
pronounced for $W'^-$ cross sections, where the uncertainties exceed
100\% at large $y_W$.

At central rapidity the $Z'$ cross section is well constrained,
with weak sensitivity to $\bar u$ and $\bar d$ quarks at the LHC
for $M_{Z'} < 3$~TeV.  The $W'$ central rapidity cross sections,
on the other hand, display sensitivity to large-$x$ PDFs of the order
$10-20\%$ for $M_{W'} \gtrsim 2$~TeV at the LHC (due to $\bar d$
and $\bar u$) and 0.75~TeV at the Tevatron (due again to $d$ quarks).
The uncertainties at central rapidity directly propagate to the
integrated cross sections, which show $\lesssim 3\%$ effects for $Z'$
production, while somewhat larger for $W'$ production, amounting to
$\lesssim 20\%$ for $W'^+$ and $\lesssim 10\%$ for $W'^-$ at the LHC
for $M_{W'} < 3$~TeV, and $\lesssim 30\%$ for $W'^+$ and $W'^-$ at
the Tevatron for $M_{W'} < 1$~TeV.

These considerations place important limits on the ability to
accurately measure heavy $W'$ and $Z'$ cross sections in hadronic
collisions, particularly at large rapidities and boson masses near
the kinematic thresholds of current colliders.  
Although our analysis is, for illustration, restricted to heavy
vector bosons with Standard Model couplings, and the quantitative
effects of the PDF uncertainties would be different in other models,
our main point is that caution must be exercised when using PDFs
in regions where these are not directly constrained, or their
uncertainties underestimated, as is the case at large $x$.
The uncertainties in the production cross sections can be reduced
by obtaining better constraints on PDFs at large $x$, especially
for the $d$ quark.
Several experiments aimed at determining the $d$ quark PDF up to
$x \approx 0.8$ are planned at Jefferson Lab following its 12~GeV
energy upgrade \cite{BoNuS12, MARATHON, SOLID}.
Uncertainties in cross sections at central rapidity, and hence in
the integrated cross sections, will also be reduced with improved
determinations of antiquark distributions at large $x$, such as
the E-906/SeaQuest experiment at Fermilab \cite{SeaQuest} which
plans to measure $\overline d/\overline u$ up to $x \approx 0.45$.

The flavor dependence of weak boson production could also be studied
with $pn$ collisions at the LHC or at the Relativistic Heavy-Ion
Collider, with the neutron provided by a beam of deuterons
\cite{Soffer94}.  Unlike for fixed target experiments, in a collider
one can study $pd$ collisions at large positive and negative rapidities.
Therefore, partons in the beam at large $x_1$ can scatter from partons
in the target at small $x_2$, and vice versa.  In particular,
measurements at large negative rapidity would be sentitive to quarks
in the deuteron at large momentum fractions, offering a probe of
nuclear corrections complementary to deuterium DIS, and constraining
the nuclear uncertainties studied in this paper.

% The flavor dependence of weak boson production could also be studied
% with $pn$ collisions at the LHC, with the neutron provided by a beam
% of deuterons \cite{Soffer94}.  The availability of a deuteron beam may
% further enhance production rates for collisions near the kinematic
% thresholds, as a result of nuclear Fermi motion, which can boost the
% momentum of a parton in a nucleus to $x > 1$ by acquiring it from other
% nucleons.  However, since deuterons at the LHC can only be accelerated
% to half of the energy of a proton, the reduced $\sqrt{s}$ in $pd$
% collisions will generally yield smaller per-nucleon cross sections
% than in $pp$.  Nonetheless, the production rates may still be amplified
% near kinematic thresholds in processes with a slow change in cross
% section with respect to $\sqrt{s}$, or with strong sensitivity to the
% high-$x$ region.  At the Relativistic Heavy-Ion Collider, on the other
% hand, the deuteron beam can be accelerated to the same energy as the
% proton beam, and studies of $pd$ collisions, at large positive and
% negative rapidities, may open new avenues for studies of nuclear
% corrections to PDFs.

Finally, while we have focussed on the $\sqrt{s} = 7$~TeV energy at
which the LHC currently operates, in future this is planned to increase
to $\sqrt{s} = 14$~TeV.  The behavior of the cross sections illustrated
here will not change qualitatively at the larger energy.  However, for
physical $W$ and $Z$ bosons, the region in rapidity where sensitivity to
nuclear models is greatest will be shifted outside of the acceptance of
current experiments, limiting the usefulness of 14~TeV data for large-$x$
PDF studies.  On the other hand, the higher center of mass energy will
increase the accessible $W', Z'$ mass range
(up to $M_{W',Z'} \approx 6$~TeV) over a larger range of rapidities.

%%%%%%%%%%%%%%%%%%%%%%%%%%%%%%%%%%%%%%%%%%%%%%%%%%%%%%%%%%%%%%%%%%%%%%%%%
\section*{Acknowledgements}

We thank D.~del~Re and J.~Erler for helpful discussions and
communications.
This work was supported by the DOE contract No.~DE-AC05-06OR23177,
under which Jefferson Science Associates, LLC operates Jefferson Lab,
DOE contract No.~DE-FG02-97ER41022, DoD's ASSURE Program, and the
National Science Foundation under NSF Contact Nos. 1062320 and 1002644.

%%%%%%%%%%%%%%%%%%%%%%%%%%%%%%%%%%%%%%%%%%%%%%%%%%%%%%%%%%%%%%%%%%%%%%%%%

\end{document}